\documentclass[twocolumn]{webofc}

\usepackage[varg]{txfonts}   
\usepackage{hyperref}
\usepackage{url}

\usepackage{kotex}
\usepackage[dvipsnames]{xcolor}

\hypersetup{colorlinks=true,citecolor=blue,urlcolor=blue,linkcolor=blue}

\begin{document}
\title {Quark-meson coupling model and heavy-ion collision}

\author{\firstname{Dae Ik} \lastname{Kim}\inst{1}\fnsep\thanks{\email{di.kim.phys@gmail.com}} \and
        \firstname{Chang-Hwan} \lastname{Lee}\inst{1}
        \and
        \firstname{Kyungil} \lastname{Kim}\inst{2}
        \and
        \firstname{Youngman} \lastname{Kim}\inst{3}
        \and
        \firstname{Sangyong} \lastname{Jeon}\inst{4}
        \and
        \firstname{Kazuo} \lastname{Tsushima}\inst{5}
}

\institute{Department of Physics, Pusan National University, Busan 46241, Korea 
\and
           Institute for Rare Isotope Science, Institute for Basic Science, Daejeon 34000, Korea
\and
           Center for Exotic Nuclear Studies, Institute for Basic Science, Daejeon 34126, Korea
\and
    Department of Physics, McGill University, Montreal H3A2T8, Quebec, Canada
\and Laboratório de Física Teórica e Computacional-LFTC, Programa de
P\'{o}sgradua\c{c}\~{a}o em Astrof\'{i}sica e F\'{i}sica Computacional,
Universidade Cidade de S\~{a}o Paulo, 01506-000 S\~{a}o Paulo, S\~{a}o Paulo, Brazil
}

\abstract{We implement the quark–meson coupling model in Daejeon Boltzmann-Uehling-Uhlenbeck (DJBUU) transport model and perform Au+Au collision simulations at intermediate energies. Results are compared with simulations using a conventional quantum hadrodynamics model. Differences in the maximum density reached during the collisions are interpreted in terms of nuclear matter properties predicted by each model.}

\maketitle
\section{Introduction}
The heavy-ion collision experiments provide the only terrestrial opportunity to investigate dense nuclear matter by momentarily creating it in the laboratory~\cite{Sorensen:2023zkk}.
One of the advantages of heavy-ion collisions is that the isospin asymmetry can be controlled by selecting the species of the projectile and target nuclei.
In particular, upcoming and ongoing rare-isotope facilities, such as RAON in Korea, RIBF in Japan, and FRIB in the United States, are expected to provide proton- or neutron-rich isotope beams, enabling studies over a broader range of isospin asymmetry.

Because dense nuclear matter cannot be observed directly, theoretical approaches such as transport models are essential for describing the full dynamical evolution of heavy-ion collisions and extracting information about dense matter from observables. To study heavy-ion collisions that will be conducted at RAON, we have developed the Daejeon Boltzmann-Uehling-Uhlenbeck (DJBUU) model~\cite{Kim:2020sjy}, named after the city where RAON is located.

As a relativistic transport model, DJBUU model allows one to study nuclear matter described by relativistic mean-field (RMF) theory with heavy-ion collisions. The quantum hadrodynamics (QHD) model, which has been commonly applied to relativistic transport models~\cite{Ko:1988zz, Song:2015hua, Buss2012Transport-theoreticalReactions}, is an RMF model based on hadronic degrees of freedom.
In contrast, the quark–meson coupling (QMC) model is an RMF model based on quark degrees of freedom. The nucleon is treated as the bag containing three quarks, which directly interact with meson mean fields. QMC model has been applied to many fields, such as nuclear matter, finite nuclei, neutron stars, and supernovae~\cite{Guichon1988,Saito1996,Guichon1996,Guichon2018,Saito1994,Stone2017}. 

In this work, we implement QMC model in DJBUU model and compare its predictions with those of QHD model. In Sec.~\ref{sec:QMC} and Sec.~\ref{sec:DJBUU}, we introduce QMC model and DJBUU model, respectively. In Sec.~\ref{sec.nmp}, we compare the results for symmetric nuclear matter and pure neutron matter obtained using QHD and QMC models. In Sec.~\ref{sec:HIC}, we perform heavy-ion collision simulations with both models and discuss the differences between the corresponding results.

\section{Quark-meson coupling model}
\label{sec:QMC}
QMC model, unlike QHD model where nucleons interact with meson fields, is a relativistic mean-field approach in which quarks confined inside a bag interact directly with meson fields.
It starts from a quark-level Lagrangian, but for practical calculations, we can use a hadron-level Lagrangian by introducing the parameter $a_N$ in the effective mass of the nucleon.

The hadron-level Lagrangian density is given by
\begin{align}
\mathcal{L} &= \bar{\psi}\big(i\gamma_{\mu}\partial^{\mu} 
             - m^*_{N}
             - g_{\omega} \gamma_\mu \omega^\mu 
             - g_{\rho} \gamma_\mu\vec{\tau}\cdot\vec{\rho}^{\mu}
             \big)\psi
\notag\\
            &+ \frac{1}{2} \bigl(\partial_\mu \sigma \partial^\mu \sigma - m_\sigma^2 \sigma^2\bigr)
            +\frac{1}{2}m_{\omega}^2\omega_{\mu}\omega^{\mu} \notag\\
            &- \frac{1}{4} \Omega_{\mu\nu} 
            \Omega^{\mu\nu} 
             +  \frac{1}{2} m_{\rho}^2 \vec{\rho}_{\mu}\cdot\vec{\rho}^\mu
           - \frac{1}{4} \vec{R}_{\mu\nu} \cdot \vec{R}^{\mu\nu} 
\label{eq:lagrangian_den}
\end{align}
where $\psi$ is a nucleon and $m^*_N$ is the effective Dirac mass. $\sigma$ is the scalar-isoscalar meson. $\omega^\mu$ and $\vec\rho^\mu$ are the vector-isoscalar and vector-isovector mesons, respectively. The vector field strength tensor for $\omega$ and $\rho$ mesons are given by
\begin{equation}
        \Omega_{\mu\nu} = \partial_\mu\omega_\nu - \partial_\nu\omega_\mu,  ~
        \vec{R}_{\mu\nu} = \partial_\mu \vec\rho_\nu - \partial_\nu \vec\rho_\mu.
\end{equation}
Here, the distinctive feature of QMC model compared to QHD model is that the effective mass includes an additional quadratic term:
\begin{equation}
    m^*_N = m_N - \big(g_\sigma\sigma - \frac{a_N}{2} (g_\sigma\sigma)^2\big).
\end{equation}
If $a_N$ equals 0, it corresponds to the effective mass in QHD model. The $\sigma$ mean-field in $m_N^*$ can be obtained by solving the equation of motion of the $\sigma$ field:
\begin{equation}
    m^2_\sigma\sigma = g_\sigma C_N(\sigma)\rho_s
\end{equation}
where $C_N(\sigma) = 1-a_N g_\sigma\sigma$. The value of $a_N = 0.181~\mathrm{fm}$ is obtained from the quark-level mean-field calculation employing a quark mass of $m_q = 5~\mathrm{MeV}$ and a bag constant of $B = (170~\mathrm{MeV})^{4}$~\cite{Tsushima:2022PTEP}.
 
Notably, recent QHD models often include nonlinear self-interaction terms of the scalar field $\sigma$ in order to reproduce a reasonable value of the incompressibility $K_0$:
\begin{equation}
U(\sigma)=\frac{1}{3}g_2\sigma^3+\frac{1}{4}g_3\sigma^4 .
\end{equation}
However, in QMC model, the parameter $a_N$ is sufficient to obtain a reasonable value of $K_0$. Therefore, we do not include $U(\sigma)$ in Eq.~(\ref{eq:lagrangian_den}) for QMC.

\section{DaeJeon Boltzmann-Uehling-Uhlenbeck model}
\label{sec:DJBUU}
DJBUU model describes heavy-ion collisions based on
the relativistic Boltzmann-Uehling-Uhlenbeck equation, which is given by
\begin{equation}
\label{eq:relbuu}
\frac{1}{E_i^{*}}\left[p^\mu \partial_\mu^x - \Bigl(p_\mu \mathcal{F}^{\mu\nu} - m_i^*   \partial_x^\nu m_i^*\Bigr)\partial_\nu^p\right] f_i(\vec{x},\vec{p}) = \mathcal{C}_i(\vec{x},\vec{p}),
\end{equation}
where $E_i^{*}=\sqrt{k^2+{m^*_N}^2}$, and the vector field strength tensor for vector field potential $V^\mu$ is $\mathcal{F}^{\mu\nu}=\partial^\mu V^\nu-\partial^\nu V^\mu$.
The left-hand side corresponds to the Vlasov equation, described by solving equations of motion of test-particles in mean-field potential, and the right-hand side $\mathcal{C}_i(\vec x ,\vec p )$ corresponds to nucleon-nucleon collision term including Pauli blocking factor.

The phase-space density is considered as the sum of test particles:
\begin{equation}
\label{eq:tpdist}
f(\vec{x}, \vec{p}) = \frac{(2\pi)^3}{N_\mathrm{TP}}
\sum_{j=1}^{A N_\mathrm{TP}} g_x\bigl(\vec{x} - \vec{x}_j\bigr)   g_p\bigl(\vec{p} - \vec{p}_j\bigr),
\end{equation}
where $A$ is the sum of mass number of projectile and target nuclei and $N_{TP}$ is the number of test particles per nucleon. $g_x$ and $g_p$ are the spatial and momentum profiles of the test particles. DJBUU uses the polynomial function which is given by
\begin{equation}
\label{eq:polyprofile}
g(\vec{u}) = \mathcal{N}  \bigl[1 - (|\vec{u}| / a_\mathrm{cut})^2\bigr]^3
\quad \text{for } 0 < |\vec{u}| / a_\mathrm{cut} < 1,
\end{equation}
where $\mathcal{N}$ is the normalization constant and $a_\mathrm{cut}$ is the cut-off parameter.

\section{Nuclear matter}
\label{sec.nmp}

To check nuclear matter properties predicted by parameters which we use in transport model, we calculate uniform nuclear matter using QHD and QMC models. For QMC model, we use the parameter set proposed in Ref.~\cite{Tsushima:2022PTEP}.
For QHD model, we use the Liu$\rho$ parameter set~\cite{Liu:2001iz}, which is commonly employed in relativistic transport models.
In addition, for comparison, we also use the NL3 parameter set~\cite{NL3-Lalazissis:1996rd}, even though it is known to predict a rather stiff equation of state~\cite{Danielewicz2002}.

We obtain the energy density from RMF theory using the energy–momentum tensor derived from the Lagrangian density. We neglect the derivative of meson fields and anti-nucleon fields. In mean-field approximation, the energy density is given by
\begin{equation}
    \varepsilon=  \sum_{i=n, p}\frac{1}{\pi^2}  \int^{k_F}_0  dk~k^2 E_i^{*}(k)+\frac{1}{2} m_\sigma^2 \sigma^2+U(\sigma)
     +\frac{1}{2} m_\omega^2 \omega^2+\frac{1}{2} m_\rho^2 \rho^2,
\end{equation}
where $E_i^{*}(k)=\sqrt{k^2+{m^*_N}^2}$, and $U(\sigma)=0$ for QMC model, as we mentioned in Sec.~\ref{sec:QMC}.

The binding energy per nucleon is defined as
\begin{equation}
    E(\rho_B,\delta) = \frac{\varepsilon(\rho_B,\delta)}{\rho_B}-M_N,
\end{equation}
where $\rho_B$ is the baryon density and $\delta=(\rho_n-\rho_p)/\rho_B$ is the isospin asymmetry. Expanding in powers of $\delta$, we obtain
\begin{equation}
    E(\rho_B,\delta) =E_0(\rho_B)+E_\mathrm{sym}(\rho_B)\delta^2+\mathcal{O}(\delta^4),
\end{equation}
where $E_0(\rho_B)$ corresponds to the binding energy per nucleon of symmetric nuclear matter and $E_\mathrm{sym}(\rho_B)$ is the symmetry energy which corresponds to the difference in the energy between pure neutron matter and symmetric nuclear matter.

Near the saturation density $\rho_0$, the quantities 
$E_{0}(\rho_B)$ and $E_{\mathrm{sym}}(\rho_B)$ 
can be written as Taylor expansions with respect to the 
dimensionless parameter $\chi = \frac{\rho_B - \rho_0}{3\rho_0}$:
\begin{equation}
    E_0(\rho_B)=E_0+\frac{1}{2}K_0\chi^2+\mathcal{O}(\chi^{3}),
\end{equation}
and
\begin{equation}
    E_\mathrm{sym}(\rho_B)=S+L\chi+\frac{1}{2}K_\mathrm{sym}\chi^2+\mathcal{O}(\chi^{3}).
\end{equation}

In this work, we calculate them using Liu$\rho$, NL3 and QMC parameter sets. The coupling constants and some nuclear matter properties, such as $E_0$, $K_0$ and $S$, are tabulated in Table~\ref{tab:constant}.

The pressure is also obtained using the energy-momentum tensor and given by
\begin{equation}
    P=  \sum_{i=n, p} \frac{1}{3\pi^2} \int^{k_F}_0  dk\frac{k^4}{E_i^{*}(k)}-\frac{1}{2} m_\sigma^2 \sigma^2-U(\sigma) 
    +\frac{1}{2} m_\omega^2 \omega^2+\frac{1}{2} m_\rho^2 \rho^2,
\end{equation}

\begin{figure}[ht]
    \centering
    \includegraphics[width=\linewidth]{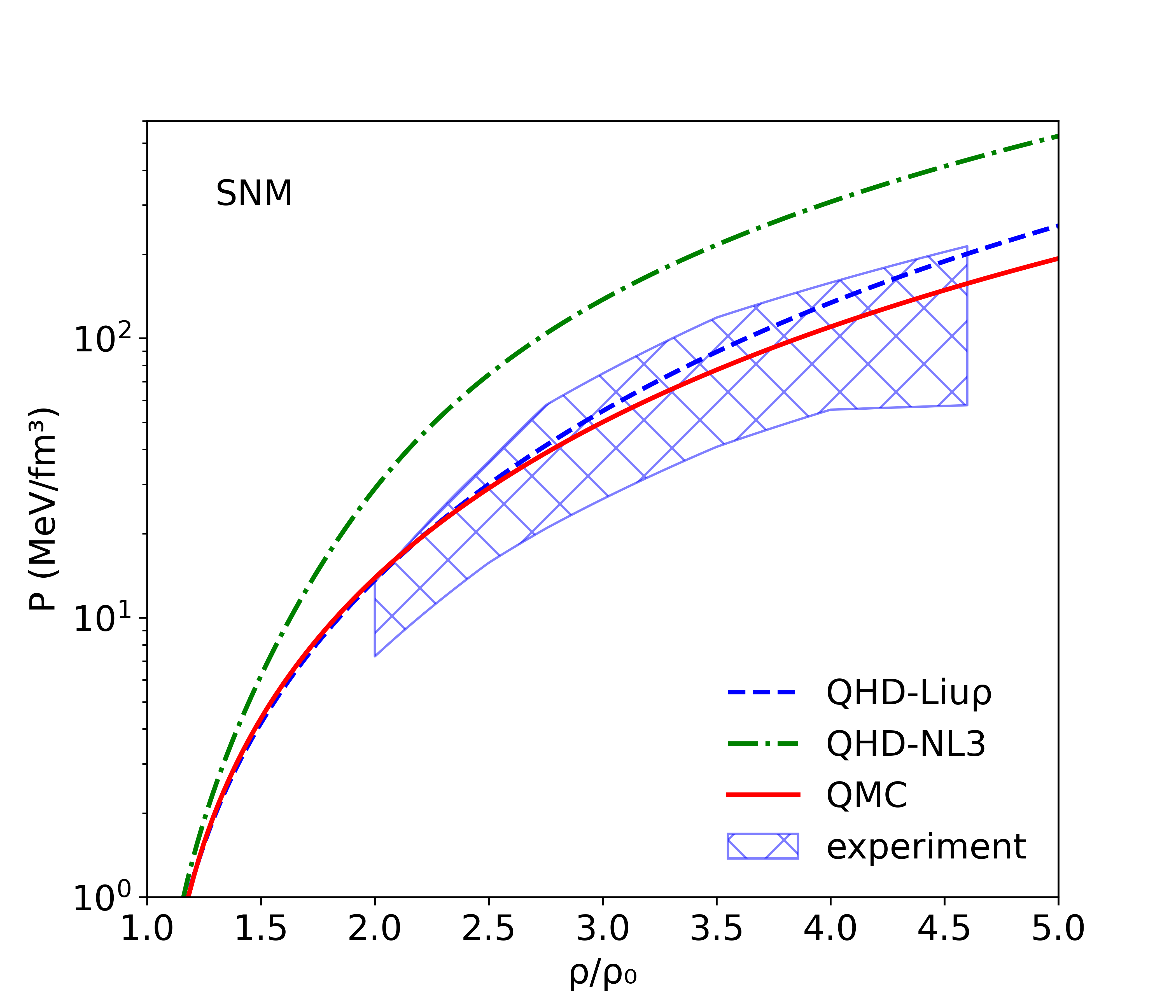}
    \caption{Pressure for symmetric nuclear matter, function of the baryon density divided by saturation density. Shaded region shows constraint from heavy-ion collisions~\cite{Danielewicz2002}.}
    \label{fig:snm}
\end{figure}

\begin{figure}[ht]
    \centering
    \includegraphics[width=\linewidth]{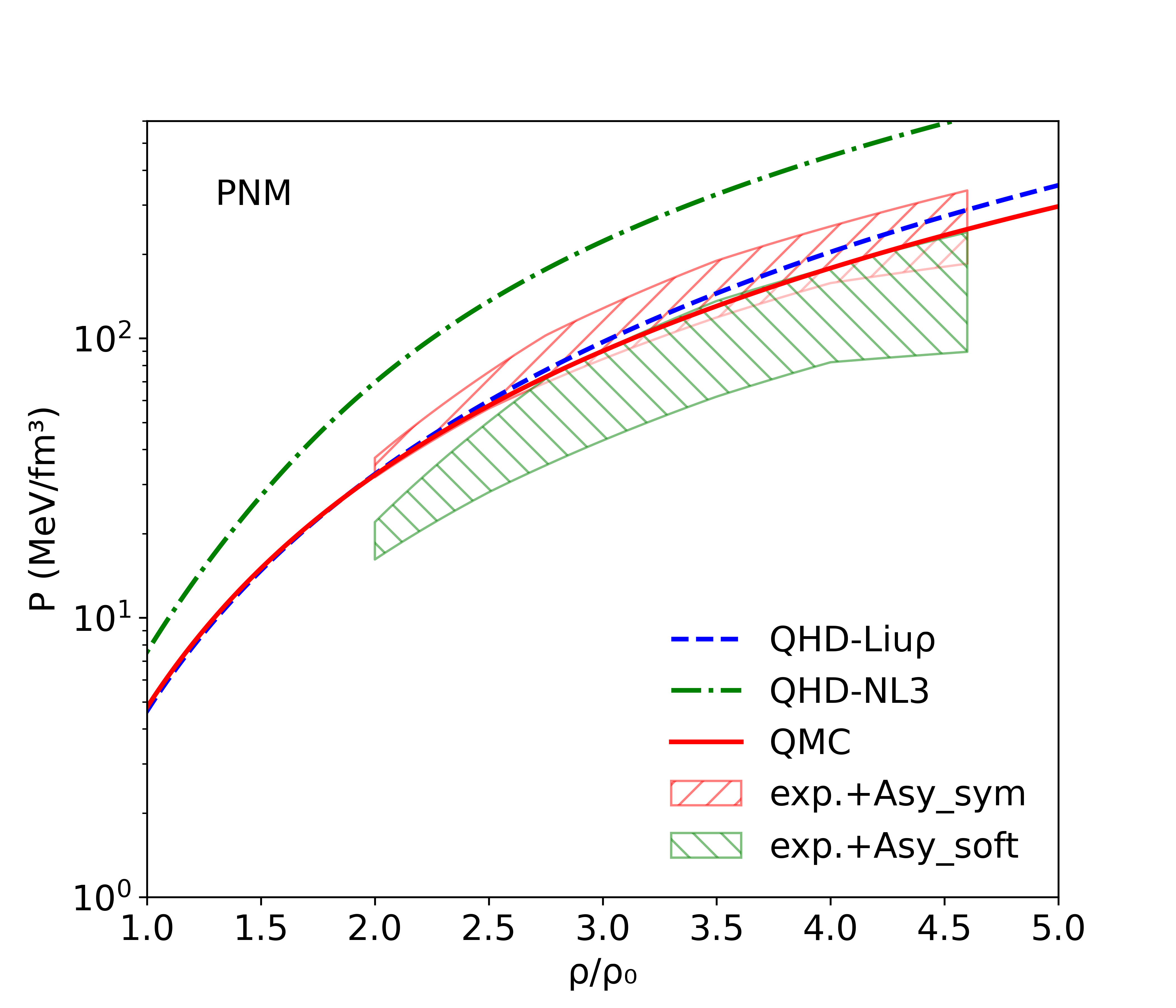}
    \caption{Pressure for pure neutron matter, function of the baryon density divided by saturation density. Shaded region shows constraint from heavy-ion collisions~\cite{Danielewicz2002}.}
    \label{fig:pnm}
\end{figure}

Figures~\ref{fig:snm} and \ref{fig:pnm} show the density-dependent pressure of symmetric nuclear matter and pure neutron matter, respectively.
When compared with the constraints from heavy-ion collision experiments~\cite{Danielewicz2002}, Liu$\rho$ and QMC lie within the constrained region, whereas the NL3 is too stiff. 

\begin{figure}[ht]
    \centering
    \includegraphics[width=\linewidth]{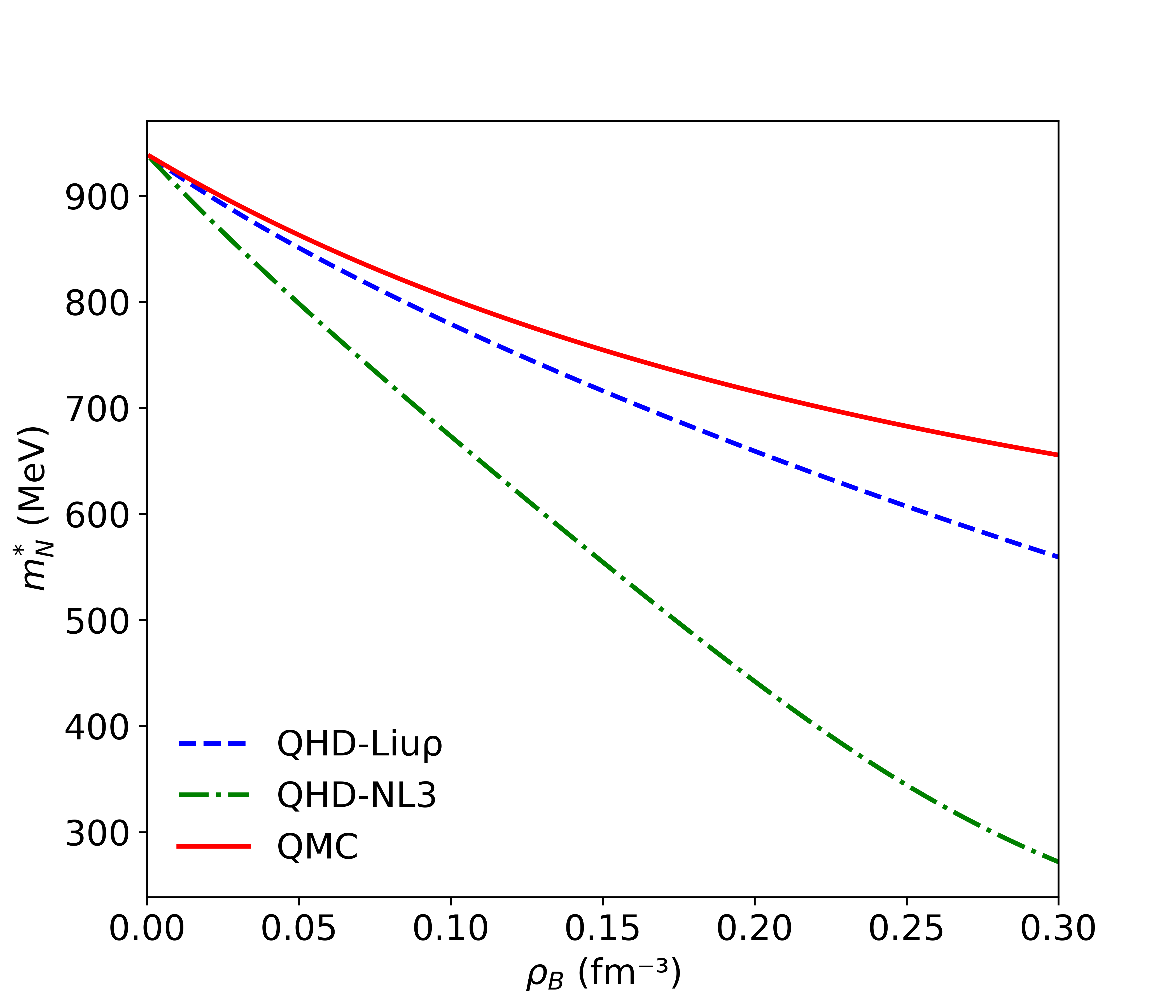}
    \caption{The effective nucleon mass, function of the baryon density.}
    \label{fig:mstar}
\end{figure}

Figure~\ref{fig:mstar} shows the density-dependent effective nucleon mass. QMC predicts the largest effective mass, followed by Liu$\rho$, while NL3 yields the smallest.
The effective masses at saturation density are tabulated in Table~\ref{tab:constant}.
 
\begin{table*}[t]
\centering
\renewcommand{\arraystretch}{1.5}
\setlength{\tabcolsep}{7pt} 
\begin{tabular}{cccccccclccc}
\hline 
 & $g_\sigma$ & $g_\omega$ & $g_\rho$ & $g_2$ & $g_3$ 
 & $a_N$ & $\rho_0$  &$m^*/m$& $E_0$ & $K_0$ & $S$ \\
 & & & & (fm)& 
 & (fm) & (fm$^{-3}$)  && (MeV) & (MeV) & (MeV) \\
\hline 
Liu$\rho$ & 8.96 & 9.24 & 3.77 & $-4.68$ & $-30.9$ & 0    & 0.16  &0.75&       $-16$&       240&       30.5\\
NL3 & 10.22 & 12.87 & 4.46 & 10.5 & $-29.0$ & 0    & 0.15  &0.6&       $-16$&       272&       37.3\\
QMC & 8.23 & 8.15 & 4.67 & 0     & 0     & 0.18 &      0.15 &0.8&       $-15.7$&       280&       35\\
\hline
\end{tabular}
\caption{Coupling constants and nuclear matter properties for the Liu$\rho$~\cite{Liu:2001iz}, NL3~\cite{NL3-Lalazissis:1996rd} and QMC~\cite{Tsushima:2022PTEP} models used in DJBUU model.}
\label{tab:constant}
\end{table*}

\section{Heavy-ion collision}
\label{sec:HIC}
In this work, we perform heavy-ion collision simulations with DJBUU model, and present results of $^{197}$Au+$^{197}$Au collision with $E_\mathrm{beam}=$ 400 A MeV and impact parameter $b$ = 4.7 fm. As mentioned in Sec.~\ref{sec.nmp}, we employ two parameter sets, Liu$\rho$~\cite{Liu:2001iz} and NL3~\cite{NL3-Lalazissis:1996rd} for QHD model and one parameter set~\cite{Tsushima:2022PTEP} for QMC model.

\begin{figure}[htbp]
    \centering
    \includegraphics[width=\linewidth]{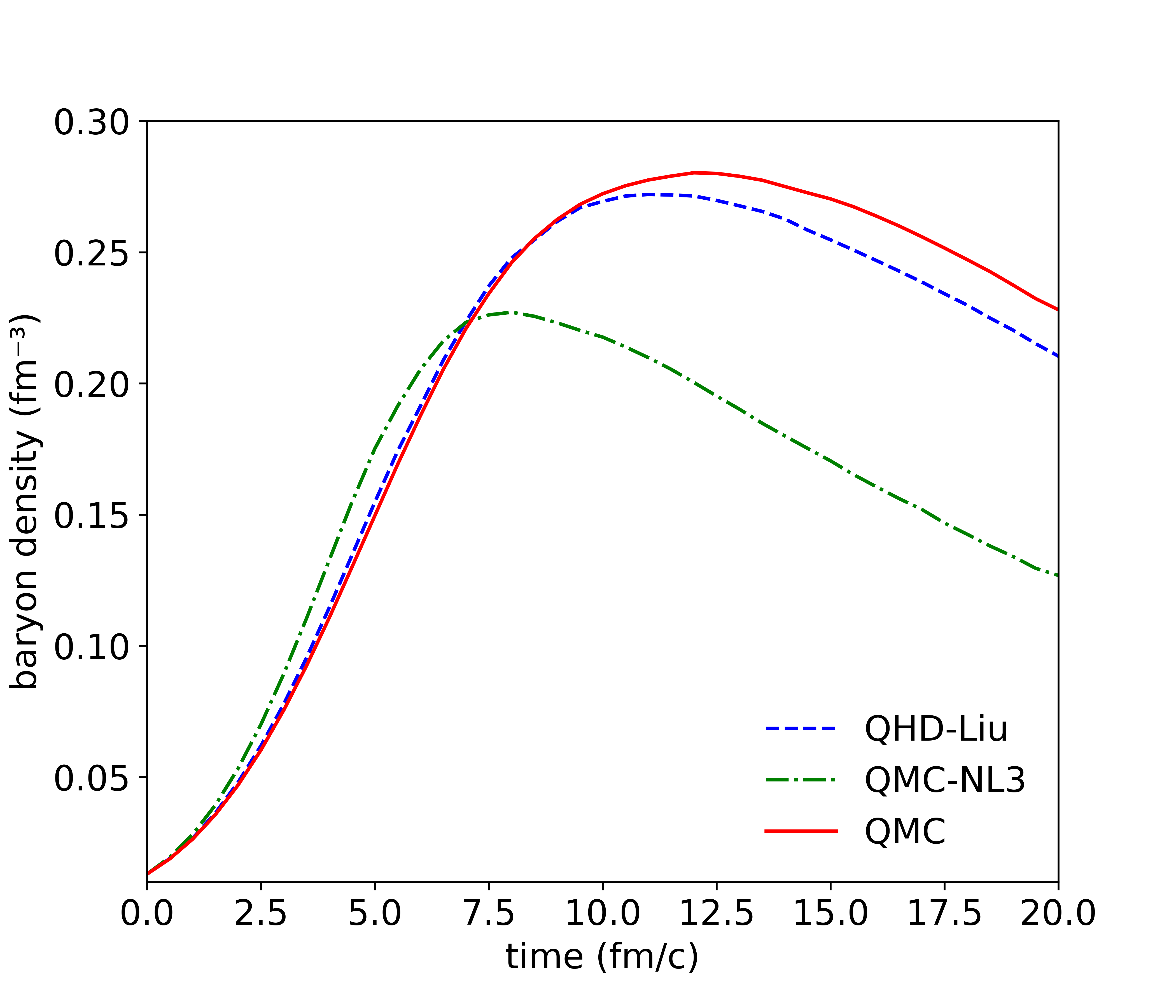}
    \caption{Time evolution of the central baryon density. The red solid line, blue dash-line and green dashed-dot line correspond to QMC, Liu$\rho$ and NL3, respectively.}
    \label{fig:density}
\end{figure}

Figure \ref{fig:density} shows the time evolution of the central baryon density at the origin in the center-of-mass frame of the system.
The central density obtained with NL3 increases more rapidly and reaches its peak earlier than the others.
Among the three parameter sets, NL3 also gives the smallest maximum density.
The results from Liu$\rho$ and QMC are quite similar up to about 8 fm/$c$; however, QMC peaks with a slightly larger maximum density than Liu$\rho$.
This behavior is consistent with the fact that the NL3 parameter set, as also shown in Figs. \ref{fig:snm} and \ref{fig:pnm}, predicts a rather stiffer equation of state, which leads to a noticeably different density evolution compared with the other two cases.

It is interesting to note that QMC yields a larger maximum density than Liu$\rho$.
We attempt to interpret this trend in terms of the nuclear matter properties listed in Table~\ref{tab:constant}.
Compared to Liu$\rho$, the larger $K_0$ and $S$ in QMC indicate a stiffer equation of state, which would generally be expected to reduce the maximum density~\cite{Long:2024ggx}.
On the other hand, the larger effective mass ratio $m^*/m$ in QMC tends to soften the equation of state and thus increase the maximum density.
If the effect of the difference in $m^*/m$ dominates over the contributions from $K_0$ and $S$, this behavior can be understood.
 
In addition, we have confirmed that the directed flow in Au+Au collisions obtained with Liu$\rho$ and QMC shows similar behavior and is consistent with experimental data. We also discuss pion production in Sn+Sn collisions; details of this analysis were presented in Ref.~\cite{Kim:2026kkf}.

\section{Summary}

In this study, we introduce QMC model, which has been successfully applied to nuclear matter, finite nuclei, and neutron stars. By implementing QMC model within DJBUU transport model, we enable heavy-ion collision simulations based on nuclear matter properties predicted by QMC model.

Before performing heavy-ion collision simulations, we examine how RMF parameter sets used in DJBUU, including the newly implemented QMC set, reproduce nuclear matter properties such as the pressure, incompressibility, and effective mass. After this verification, we simulate Au+Au collisions using QHD parameter sets (Liu$\rho$ and NL3) together with QMC set, focusing on the time evolution of the central density.

We find that NL3, which exhibits a stiffer behavior in nuclear matter calculations, leads to the lowest maximum density and shows a markedly different trend from the other parameter sets. Liu$\rho$ and QMC sets yield similar density evolutions, although QMC produces a slightly smaller maximum density. Finally, we discuss the difference between Liu$\rho$ and QMC in relation to nuclear matter properties such as the incompressibility, symmetry energy, and effective mass.\\

\section{Acknowledgments}
D.I.K and C.-H.L were supported by the National Research Foundation of Korea (NRF) grant funded by the Korean government (No. RS-2023-NR076639).
D.I.K was supported by the Hyundai Motor Chung Mong-Koo Foundation and by the 2023 BK21 FOUR Graduate School Innovation
Support funded by the Pusan National University (PNU-Fellowship Program).
K.K. and Y.K. were supported in part by the Institute for Basic Science (2013M7A1A1075764, IBS-I001-01, IBS-R031-D1).
S.J.~acknowledges the support of the Natural Sciences and Engineering Research Council of Canada (NSERC) [SAPIN-2024-00026].
K.T.~was supported by Conselho Nacional de
Desenvolvimento Cient\'{i}fico e Tecnol\'ogico (CNPq, Brazil), Processes No. 304199/2022-2, and
FAPESP Process No.~2023/07313-6, and his work was also part of the projects, Instituto Nacional de
Ci\^{e}ncia e Tecnologia - Nuclear Physics and Applications (INCT-FNA), Brazil, Process
No.~464898/2014-5.
\\

\bibliographystyle{aip}
\bibliography{biblio}

\end{document}